\documentstyle[prd,aps,floats,twocolumn,tighten]{revtex}

\def\lsim{\mathrel{\rlap{\lower4pt\hbox{\hskip1pt$\sim$}}
    \raise1pt\hbox{$<$}}}         
\def\gsim{\mathrel{\rlap{\lower4pt\hbox{\hskip1pt$\sim$}}
    \raise1pt\hbox{$>$}}}         

\newcommand{\ApJ}[3]{Astrophys. J.\ {\bf #1}, #3 (#2)}
\newcommand{\mnras}[3]{Mon. Not. R. Astron. Soc.\ {\bf #1}, #3 (#2)}

\newcommand{\physrevd}[3]{Phys. Rev. D\ {\bf #1}, #3 (#2)}

\newcommand{\physrevlett}[3]{Phys. Rev. Lett.\ {\bf #1}, #3 (#2)}
\newcommand{\hepph}[1]{hep-ph/#1}
\newcommand{\astroph}[1]{astro-ph/#1}
\def\beq{\begin{equation}}
\def\eeq{\end{equation}}
\def\bey{\begin{eqnarray}}
\def\eey{\end{eqnarray}}

\def\msun{\;{\rm M}_\odot}

\def\spose#1{\hbox to 0pt{#1\hss}}
\def\lta{\mathrel{\spose{\lower 3pt\hbox{$\sim$}}
    \raise 2.0pt\hbox{$<$}}}
\def\gta{\mathrel{\spose{\lower 3pt\hbox{$\sim$}}
    \raise 2.0pt\hbox{$>$}}}

\long\def\comment#1{}

\input{psfig.tex}
\begin{document}
\draft
\twocolumn[\hsize\textwidth\columnwidth\hsize\csname @twocolumnfalse\endcsname
\title{A Dark-Matter Spike at the Galactic Center?}
\author{Piero Ullio$^{1,2}$, HongSheng
Zhao$^{3}$, and Marc Kamionkowski$^1$}
\address{$^1$California Institute of Technology, Mail Code 130-33,
Pasadena, CA 91125.  Email: {\tt kamion@tapir.caltech.edu}}
\address{$2$SISSA, via Beirut 4, 34014 Trieste, Italy.  Email:
{\tt ullio@he.sissa.it}}
\address{$^3$Institute of Astronomy, Madingley Road, Cambridge, CB30HA, UK.  
Email: {\tt hsz@ast.cam.ac.uk}}

\date{January 2001}
\maketitle

\begin{abstract}
The past growth of the central black hole (BH) might have enhanced the
density of cold dark matter halo particles at the Galactic center.  We
compute this effect in realistic growth models of the present
$(2-3)\times 10^6 M_\odot$ BH from a low-mass seed BH, with special
attention to dynamical modeling in a realistic galaxy environment with
merger and orbital decay of a seed BH formed generally outside the
exact center of the halo.  An intriguing ``very-dense spike'' of dark
matter has been claimed in models of Gondolo and Silk with density
high enough to contradict with experimental upper bounds of neutralino
annihilation radiation.  This ``spike'' disappears completely or is
greatly weakened when we include important dynamical processes neglected in
their idealized/restrictive picture with cold particles surrounding an
at-the-center zero-seed adiabaticly-growing BH.  For the seed BH to
spiral in and settle to the center within a Hubble time by dynamical
friction, the seed mass must be at least a significant fraction of the
present BH.  Any subsequent at-the-center growth of the BH and
steepening of the central Keplerian potential well can squeeze the
halo density distribution only mildly, whether the squeezing happens
adiabatically or instantaneously.
\end{abstract}

]

\narrowtext

\section{Introduction}

Weakly interacting massive particles (WIMPs) are among the leading 
candidates for the dark matter in galactic halos.
Such particles arise in extensions to the standard model
of particle physics.  The most widely studied example is the
neutralino, plausibly the lightest superpartner in
supersymmetric versions of the standard model.  The cosmological
abundance of WIMPs is naturally of order the critical density of
the Universe, and so WIMPs make natural dark-matter candidates.

The search for WIMPs in the Milky Way halo has been a major endeavor 
both theoretically and experimentally in the last two decades
(for comprehensive reviews see Refs.~\cite{jkg,lars}).
The main efforts so far have been to detect WIMPs directly in 
low-background laboratory experiments and indirectly via observation 
of the energetic neutrinos produced by annihilation of WIMPs that have
accreted in the Sun and/or Earth. Another possibility is to search
for exotic cosmic rays, such as high-energy gamma rays,
positrons, antiprotons, or neutrinos produced by WIMP pair annihilations in
the Galactic halo.   In particular, the flux of gamma rays in a
given direction is proportional to the line-of-sight integral of
the square of the WIMP density.  Since the dark-matter density
is expected to be largest towards the Galactic center, the
flux of such exotic gamma rays should be highest in that direction.

We now know that black holes of mass of $10^6-10^8\msun$
reside in most galactic centers (see, e.g., \cite{HoKor00}).  
In particular, the arguments
for a $2.6\times 10^6 \msun$ black hole at the center of the Milky
Way are at this point almost incontrovertible (see, e.g., 
Refs.~\cite{Genzeletal}) given the rise of the velocity 
(radial and proper motion) dispersion within 1 pc of the
Galactic center. It has long been argued that if a black hole
grows inside a given population of stars, the density of stars
should be enhanced in the vicinity of the black hole after
its appearance \cite{peebles,young,ipser,qhs}.  
If the WIMP density is likewise enhanced at the Galactic center,
then the source of exotic cosmic rays from WIMP annihilation
therein should be increased.  Several authors investigated the 
enhancement in the gamma-ray flux for the dark-matter-density
spike, $\rho(r) \propto r^{-1.5}$ as the radius $r\rightarrow 0$, 
that would be induced by the adiabatic growth of a black hole in 
an initially isothermal dark-matter profile \cite{ipser}.

More recently, Gondolo and Silk~\cite{GS} (hereafter GS)
investigated the dark-matter spike that arises if a black hole
grows adiabatically at the center of a dark-matter halo that
initially has a singular power-law cusp, $\rho(r) \propto
r^{-\gamma}$ with $0<\gamma<2$, as suggested by numerical
simulations \cite{NFW,moore}.
They found that the black-hole growth gives rise to a dark-matter
spike $\rho(r) \propto r^{-A}$ with $2.25\leq A \leq 2.5$, in
agreement with earlier scaling results found in Ref. \cite{qhs}.
In this case, the cosmic-ray source from WIMP pair annihilation, 
proportional to $\int \, d^3r\, \rho^2(r)$, becomes huge due to the 
diverging contribution from small radii.  (The divergence is cut 
off by the black-hole horizon or by a maximum density  
set by WIMP pair annihilation.)  The flux
of neutrinos from neutralino annihilation in this spike exceeds 
current experimental upper bounds in some regions of neutralino
parameter space.  Subsequent work \cite{gondolo,BSS} shows
that synchrotron emission from the motion of electrons and
positrons produced by 
neutralino annihilation in this spike would also be significant
and in most cases exceeds current upper bounds.  The conclusion
of Refs. \cite{GS,gondolo,BSS} is that either neutralino dark matter
or a cuspy halo must be ruled out.

Given these very important implications for dark-matter searches, 
the claim of the possible presence of a steep dark-matter spike
at the Galactic center and the underlying assumptions should be 
investigated more carefully.  Here we argue that although a possibility, 
the enhancement found by GS requires somewhat unusual initial conditions 
for the Galactic halo and for the black hole. In particular, we emphasize
how crucial it is that the black hole grew adiabatically
from a tiny initial mass to its present-day mass, and that this happened
precisely at the center of the dark-matter distribution.

To do so, we consider a few 
alternatives for the black-hole-formation scenario that we can 
treat with semi-analytic methods and jet captures the main dynamical processes 
in (hierarachical) galaxy formation.  We show that for the GS spike 
to arise, the majority of the black-hole growth had to occur
within the inner 50 pc of the dark-matter cusp and on a
timescale long compared with $10^7$ years.  We also show that
the growth of the spike relies on the existence of a very high
density of cold orbits within the inner 50 pc.  If the black hole is 
formed away from the center and then spirals in, it should
generically lead to the weaker dark-matter spike, $\rho \propto
r^{-0.5}$, with possibly even a depletion, rather than an
enhancement, in the central dark-matter density.

The outline of the paper is as follows: In Section~\ref{warmup},
we consider a simple toy model to illustrate the origin of the spike 
found by GS in case the black hole grows adiabatically at the exact center 
of the dark-matter system. In Section~\ref{sec:adia} we perform the
computation of the adiabatic growth for a class of more realistic
dark-matter-halo profiles, stressing a few more subtle points.
In Section~\ref{sec:inst} we treat instead the opposite case of 
fast growth for the black hole. Section~\ref{sec:spiral} deals with 
the more complex scenarios of off-centered formation for the black hole.
While in all cases above we include just dark matter and the black 
hole, in Section~\ref{sec:stars} we consider the more realistic
scenario in which other baryonic components (i.e., stars in the
stellar nucleus, bulge, and disk) are included as well.
In Section~\ref{sec:concl} we summarize and comment on current
theoretical ideas about the origin of the black hole.

\section{Warmup: Halo distributions with only circular orbits}
\label{warmup}

Before proceeding with our numerical calculations, let us
consider a simple toy model that illustrates the origin of the GS
spike.  In this model, we assume an initial self-gravitating 
dark-matter distribution $\rho_i(r)$ made of particles
that are all on circular orbits, at the center of which a black
hole grows adiabatically.  
The angular momentum of each particle remains invariant
since the central black hole 
exerts no  torque on any dark-matter particle.  This tells
us that  $r_i M_i(r_i)= r_f M_f(r_f)$, where $M(r)$ is the total
(i.e., dark-matter plus black-hole) mass enclosed within the
radius $r$.  Moreover, conservation of the dark-matter mass tells us
that $M_i^h(r_i)= M_f^h(r_f)$, where the superscript $h$ denotes 
the halo contribution to the mass.  

Now suppose the initial dark-matter density profile is
$\rho_i(r_i) \propto r_i^{-\gamma}$ and let the final density
be $\rho_f(r_f) \propto r_f^{-A}$. Conservation of mass
implies that $r_i^{3-\gamma} \propto r_f^{3-A}$.  
In the final state, the mass enclosed within a radius $r$ will
be dominated by the black hole, $M_f(r_f) \simeq M_{\rm BH}$, as 
$r\rightarrow0$.  Conservation of angular momentum
thus gives $r_i^{4-\gamma} \propto r_f$.  Putting these two
conditions together, we find a final profile, $\rho_f(r_f)
\propto r_f^{-A}$, where $A=(9-2\gamma)/(4-\gamma)$
\cite{qhs,GS}.  Note that for $0<\gamma<2$, we get $2.25<A<2.5$.
Hence the WIMP pair annihilation rate, which is proportional to
the density $\rho_f(r_f)$ squared, would have diverged except for
the truncation at the horizon of the BH.

This derivation illustrates several important points, some of
which will be discussed further for more general distributions
in subsequent Sections.

\subsection{Where does the mass in the spike come from?}  

The dark-matter particles that make up the spike come
from the inner part of the dark-matter halo, the part that encloses 
a mass roughly equal to $M_{\rm BH} \simeq 2.6 \times 10^6 M_\odot$. 
The addition of a black hole will have negligible effect on circular 
orbits that enclose larger masses.  
\comment{For a more general (and plausible)
phase-space distribution, it is more difficult to show analytically that }
That the spike comes from this inner part agrees with intuitions, 
and is suggested by numerical simulation of
Ref. \cite{qhs} (see their Fig. 6) 
for models with isotropic velocity distributions.

For the power-law density profile that GS considered (and for
the NFW profile we consider below), the radius that
encloses $2.6 \times 10^6 \, M_\odot$ of halo mass is only about 50-60
pc.  Thus, for the spike in the GS picture to form, the black 
hole must have originated within the inner 50--60 pc of the Galactic 
center.  This is in some
sense surprising because this length scale is
microscopic compared with characteristic Galactic distance
scales: $\sim3$ kpc for the bulge and $\sim10$ kpc for the disk.  Put another
way, the inner 50 pc encloses only about $(1.5-2) \times 10^8 \, M_\odot$ 
of stellar material, about $5 \times 10^{-3}$ of the total stellar mass
(probably about $(5-8) \times 10^{10}\, M_\odot$) of the
galaxy.  Although the alignment between the black hole and the
stellar distribution is quite good in our own Galaxy 
(even that is a mystery---the situation is quite different, e.g., 
in M31), the required alignment is between the black hole and
the center of the {\it dark-matter} system.
Although baryons and dark matter are mixed homogeneously in the
early Universe, there are numerous processes involved in galaxy
formation that should segregate the baryons from the dark
matter.  For example, baryons will experience radiation drag
from the cosmic microwave background when they first start to
cool; UV radiation from the first generation of stars could
exert a wind on the baryons; in galaxy mergers, baryons shock
while dark-matter halos pass through each other; baryons today
appear in disks, bulges, and bars, unlike dark matter; they are
clumped into globular clusters, open clusters, and molecular
clouds.  Thus, there are many good reasons to believe that the
precise center of the baryon distribution should be displaced
relative to that of the dark matter, at least at some intermediate
stages in the formation of our Galaxy.  Thus, even if the black
hole formed at the center of a stellar distribution, it would
still be somewhat surprising if, at all times during its formation, 
that coincided to 50 pc with the center of the dark-matter distribution.

\subsection{Cold particles}

The circular-orbit toy model shows us that the particles that
give rise to the spike are initially very cold; i.e., the
circular speeds in the initial distribution approach zero as
$r\rightarrow0$.  (Below, we will show that the same is true for
more general and realistic velocity distributions.)
Again, this is somewhat surprising, as cold
particles are easily subject to heating by a variety of
phenomena, especially in the  environment at the
Galactic center.  For example, the passing of molecular clouds, open,
or globular clusters would tend to disrupt these cold orbits, as 
would the existence of a bar.  Likewise, the mergers and
sub-halo accretion that certainly played a role in the formation 
of the Milky Way would tend to disrupt these cold orbits.
We cannot be more quantitative without a more precise
model for galaxy formation.  However, these disruptive processes 
must have occurred at some level, and they would tend to hinder
the growth of the dark-matter spike.

Note that a dark-matter spike that
will give rise to a divergent WIMP annihilation rate can arise even
if $\gamma=0$.  As explained in Ref. \cite{qhs}, implied above,
and discussed further below, the existence of a spike is more
generally due to the existence of a singularity in the density
of cold orbits at the Galactic center.  Such singularities can
arise even if the density profile is non-singular.  
If the density $\rho$ at the origin is finite
and can be written as an analytic function of the Cartesian
coordinates $x$, $y$, and $z$, then there will be no singularity 
in the phase-space distribution that gives rise to a spike, but
otherwise there will be a phase-space singularity and a spike
after adiabatic growth of a black hole \cite{qhs}.  
Thus, consider for example, the profiles $\rho(r) \propto
(a+r)^{-2}$ and $\rho(r) \propto (a^2+r^2)^{-1}$, both of which
approach a constant at the  origin, and suppose that both have
isotropic velocity distributions.  As argued in Ref. \cite{qhs}
and described below, adiabatic black-hole growth
in the former gives rise to a spike $\rho\propto r^{-A}$ with
$A=2$, while the latter gives rise to a spike with $A=1.5$.

\subsection{Initial black-hole mass}

The circular-orbit model also
indicates that the spike will grow
only if the black hole grows almost entirely (not just
partially) at the black-hole center.  Put another way, the spike 
that grows from an initial state with a $10^6\,M_\odot$ black
hole and a $\rho\propto r^{-1}$ dark-matter halo to a final
state with a $2\times10^6\, M_\odot$ black hole is not nearly as 
dramatic as that which arises when a black hole is grown from
nothing to $10^6 \, M_\odot$.  If we start with a black hole of
significant mass and a dark-matter cusp, then the angular
momentum of the inner orbits is dominated by the initial
black-hole mass.  Slowly doubling the black-hole mass reduces
the radius of each of these inner orbits by a factor of 2 and
thus leads to a factor-of-4 density enhancement.  In other
words, the huge densities of the innermost orbits that give rise 
to a huge annihilation signal come from the fact that initially, 
$M_i(r)\rightarrow0$ as $r\rightarrow0$ and the radius $r_i$ of each
inner orbit is reduced by a factor $\sim M_{\rm BH}/M_i(r_i)$.  
If a black hole exists at the Galactic center
 from the start with a mass $M_{\rm BH}^i$ and then grows to a
final mass $M_{\rm BH}^f$, the initial dark-matter density profile
$\rho_i(r_i) \propto r_i^{-\gamma}$ becomes a density profile
with the same exponent and normalization enhanced by a factor
$\sim (M_{\rm BH}^f/M_{\rm BH}^i)^{3-\gamma}$. 

\subsection{Timescale for adiabatic growth}

If the initial halo profile is $\rho(r)\sim r^{-1}$, then the
enclosed mass will be $M(r) \propto r^2$, the circular velocity
will be $v_c(r)\propto r^{1/2}$ and the period of a circular
orbit will be $T\propto r^{1/2}$.  The period of an orbit
at 50 pc will therefore be $\sim10^{7}$ yr.  
Thus, the black hole can grow relatively
quickly compared with a cosmological timescale and still be
considered adiabatically.  This provides an argument in {\it
favor} of the GS spike.  On the other hand, the present-day
growth rate of the black hole due to consumption of stars coming
inside its loss cone is at most $3\times 10^4 \msun$/Gyr
\cite{MT}, one order of magnitude smaller than what is needed to
form the current black hole in a Hubble time (mean growth rate
of $2\times 10^5 \msun$/Gyr).  This information combined with
the evidence that peak AGN activity occurred near redshift
$z\sim2$ suggests that the majority of the black-hole mass was
in place at these early epochs.  Although this leaves plenty of
time (few Gyr) for the black hole to grow adiabatically, the
violent processes that are likely required to form the black
holes that power AGN may well operate on a much more rapid
timescale.

\subsection{Weaker spikes at smooth densities}

All of the discussion above should make it clear that if the
black hole grows adiabatically away from the Galactic center, or 
in a Galactic center with a nonsingular phase-space
distribution, then the spike should be the weaker $\rho\propto
r^{-1.5}$ spike found in Refs. \cite{peebles,young,ipser,qhs}.  Although this 
will produce an enhancement in the flux of annihilation
radiation, it will not produce the huge enhancement that arises
in a spike $\rho(r) \propto r^{-A}$ with $A>2$.

\section{Adiabatic growth of the black hole}
\label{sec:adia}

We now consider adiabatic growth of a black hole at the center
of a spherically symmetric self-gravitating dark-matter system
with a more general (and realistic) phase-space distribution.
Because of the two key ingredients here---adiabaticity and spherical 
symmetry---this problem can be treated semi-analytically.   
The slow changes in the potential allow us to 
relate in a simple way the dark-matter distribution functions before 
and after the appearance of the black hole. The symmetry, both
in the initial and final states, implies simple forms for
the adiabatic invariants. 

The calculation has been discussed in several 
papers in the past, in the context of growth of a black hole in a
general stellar system \cite{peebles,young,ipser,qhs}, or in the
context of growth of the central black hole at the Milky Way
center \cite{GS} (see also Ref.~\cite{ipser}). 
Below we briefly review the calculation, both
to reproduce the result in GS and to provide some additional
insight on the origin of the spike.

\subsection{Semi-analytic treatment}
\label{sec:anal}

In a spherically symmetric problem, the adiabatic invariants 
are the angular momentum and the radial action; i.e., $L_f = L_i
\equiv L$ and $J_{r,f}(E_f,L)=J_{r,i}(E_i,L)$, where~\cite{bt}
\begin{eqnarray}
     J_r(E,L) & = &\frac{1}{\pi} \int_{r_{\rm min}}^{r_{\rm max}} 
     dr\,v_r \nonumber \\
     & = &\frac{1}{\pi} \int_{r_{\rm min}}^{r_{\rm max}} 
     dr\,\sqrt{2(E-\Phi(r))-\frac{L^2}{r^2}}  ,
\end{eqnarray}
and the subscripts $i$ and $f$ refer to the value of the
quantity in the initial and final state. If the growth of the
black hole is adiabatic, the distribution function $f(E,L)$ remains
unchanged, so $f_f(E_f,L) = f_i(E_i,L)$.

We specify the initial state by its density $\rho_i(r)$ as
a function of radius $r$.  We also postulate that it has an 
isotropic velocity-dispersion tensor.  If so, then there is a
one-to-one correspondence between $\rho_i(r)$ and $f_i(E_i)$,
given by Eddington's formula (see, e.g., \cite{bt}),
\begin{equation}
     f(E) =  \frac{1}{\sqrt{8} \pi^2}
     \left[ \int_{E}^0 \frac{d^2\rho}{d\Phi^2}
     \frac{d\Phi}{\sqrt{\Phi-E}} 
     - \frac{\left({d\rho}/{d\Phi}\right)_{\Phi=0}}
     {\sqrt{-E}}\right]\,,
\label{eq:edd}
\end{equation}
and the initial state is fully defined.

By inverting the expression for the radial action in the initial state, 
we can write $E_i$ as a function of $E_f$ and $L$ and hence find
$f_f(E_f,L)$ from $f_i(E_i,L)$. To compute the radial action in
the final state, we actually need first a guess for the final
potential.  We assume that the potential is dominated by the
the black hole in its vicinity, while it is roughly unchanged
at large galactocentric distances, fixing 
$\Phi_f(r) \cong \Phi_i(r)- G M_{\rm BH}/r$.  
Once $f_f(E_f,L_f)$ is obtained, the density profile of the 
system in the final state is recovered by integrating $f_f$ over 
the velocity space, or, with a change of variables, over
$E_f$ and $L$:
\begin{equation}
     \rho_f(r) = \frac{4 \pi}{r^2} \int_{E_f^{\rm min}}^{0} dE_f 
     \int_{L_f^{\rm min}}^{L_f^{\rm max}} dL_f \frac{L_f}{v_r} f_f(E_f,L_f)\,.
\label{eq:finprof}
\end{equation}
Because of the form we choose for $\Phi_f(r)$, we expect
$\rho_f(r)$ to be fairly accurate in the central region and far
 from the black hole. A more accurate result could be
obtained in the transition region by iterating. However, 
we will mainly be interested in the region near the black hole 
where the approximation $\Phi_f(r) \simeq -G M_{\rm BH}/r$ is fairly 
accurate.  It is also the form adopted by GS.

\begin{figure}[t]
\vskip -1.0cm
\centerline{\psfig{file=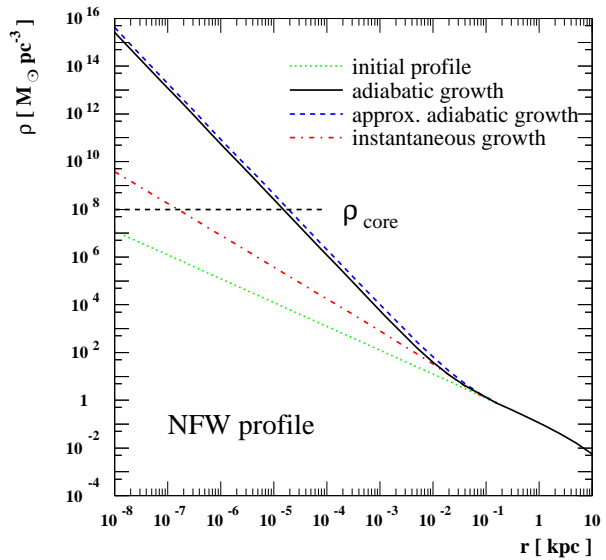,width=3.5in}}
\caption{Enhancement of an NFW dark-matter-density profile due to the
growth of the Galactic black hole at the center of the dark-matter
system. The initial profile (dotted curve) is modified into the solid
curve if the growth of the black hole is adiabatic, or into the 
dash-dot curve for a sudden growth. Also shown is the case of
adiabatic growth for an initial profile with all particles on circular
orbits (dashed curve). $\rho_{\rm{core}}$ is the maximum WIMP density
above which WIMPs are depleted by pair annihilations.} 
\vskip -0.2cm
\label{fig:fig1}
\end{figure}

We apply this procedure numerically in one case. Among the general 
family of spherical double-power-law dark-matter profiles (the
Zhao models \cite{hsz}),
\begin{equation}
     \rho_{i}(r) = \rho_0 \left(\frac{R_0}{r}\right)^{\gamma}
     \left[\frac{1+(R_0/a)^{\alpha}}{1+(r/a)^{\alpha}}\right]^
     {(\beta-\gamma)/\alpha},
\label{eq:prof1}
\end{equation}
we focus on the functional form suggested by the N-body simulation
of Navarro, Frenk, and White~\cite{NFW}, a density profile 
with $(\alpha,\beta,\gamma)=(1,3,1)$ (hereafter the NFW profile). 
We fix the values of the Sun's galactocentric distance, the local 
halo density and the length scale in the profile, respectively, 
as $R_0 = 8$~kpc, $\rho_0 = 0.3$~GeV~cm$^{-3}$ and $a = 20$~kpc 
(values compatible with available dynamical constraints); 
this profile is plotted as a dotted line in Fig.~\ref{fig:fig1}. 

We first compute numerically
the corresponding distribution function and then we follow the same 
steps applied in Ref.~\cite{young} for the isothermal sphere.
We perform numerically the inversion between $E_i$ and $E_f$,
computing the radial action for the initial and the final state
on grids, respectively, in the $(E_i,L)$ and $(E_f,L)$ plane 
(we implement adaptive grids in order to achieve everywhere 
the required numerical accuracy for the inversion). 
Then we compute $\rho_f(r)$ performing the double integral
in Eq.~(\ref{eq:finprof}). The result is shown as a solid 
line in Fig.~\ref{fig:fig1}. As can be seen, 
the adiabatic growth of the black hole at the center of the
dark-matter system induces a sharp increase in the dark-matter
density close to the black hole (compare the solid line with the
dotted  line which gives the initial profile).
At a galactocentric distance of about 20 pc, there is the transition
 from the $1/r$ cusp in the NFW profile to a much steeper spike
$\rho(r) \propto r^{-A}$ with $A=7/3$.  This is in agreement
with the corresponding result 
(profile scaling as $1/r^{\gamma}$ with $\gamma =1$) in GS.
In Fig.~\ref{fig:fig1} we also plot, as a dashed line, the result
for the adiabatic contraction if all particles are on circular
orbits (the toy model we introduced in the previous Section).
As can be seen, this simplified treatment reproduces quite
closely the result we obtained in case of an isotropic 
distribution (within a factor of 2, which is less than the uncertainty
in the overall normalization of the dark-matter profile).

The density in the spike exceeds $10^8 \msun$~pc$^{-3}$,
which is roughly the maximum density if the dark matter is
made of WIMPs, which we indicate as $\rho_{\rm{core}}$
in this and in the following Figures. The limit comes from 
the value of the pair-annihilation cross section for WIMPs, which 
is fixed, to some extent, by their relic abundance (see GS 
for details).  In fact, the cross section for WIMP-WIMP elastic
scattering should generally be larger than the annihilation
cross section, so the maximum central density derived by GS
should actually be a bit lower.

\subsection{Scaling properties}

The result for the spike slope we have just derived numerically 
could have been anticipated 
using scaling properties, as suggested in Ref.~\cite{qhs}. 
This argument can be summarized as follows: 
Let the potential of the initial system increase with radius,
for small radii, as
$\Phi_i(r) \sim r^{2-\gamma}$; this is the leading term in systems
such as, e.g., those described by Eq.~(\ref{eq:prof1}) with
$\gamma \geq 0$.  Then the conservation of the radial action implies
the scaling $E_i \sim E_f^{-(2-\gamma)/(4-\gamma)}$~\cite{qhs}.   
Suppose then that the distribution function in the initial state 
diverges near $E_i = \Phi_i(0)$ as $f_i(E_i) \sim
[E_i - \Phi_i(0)]^{-n}$. Combining this with the expression for the 
final density profile, Eq.~(\ref{eq:finprof}), it follows that
\begin{equation} 
     \rho_f(r) \sim r^{-A}\,,\;\;\;\;\;A=\frac{3}{2}
     +n\,\left(\frac{2-\gamma}{4-\gamma}\right)\,.
\label{eq:scal}	
\end{equation}

It is interesting to derive the exponents $A$ and $n$ for the
family of profiles introduced above in Eq.~(\ref{eq:prof1}).
We derive $n$ for $1 \leq \alpha \leq 2$;
results for other ranges can be obtained analogously. In
Eddington's formula, Eq.~(\ref{eq:edd}),
the second term is zero, while the main contribution to the first
comes from the singularity in the lower extreme of integration.
After some algebra, one finds that the leading terms are given by
\begin{equation}
     f(E) \sim \int_{r^{\star}} dr 
     \frac{\left[c_0 + c_1 (r/a)^{\alpha} + c_2 (r/a)^{2\alpha} 
     + \ldots \right]}
     {r^3\;\sqrt{r^{2-\gamma}-[E - \Phi(0)]}},
\end{equation}
where $r^{\star} = [E - \Phi(0)]^{1/(2-\gamma)}$, and the first
two coefficients in the expansion are (up to a constant),
\begin{eqnarray}
     c_0 & = & -2 \gamma (3-\gamma) \nonumber \\
     c_1 & = & (\alpha-2)(3-\gamma)(\beta-\gamma) \nonumber \\
     && \times \frac{(2\alpha-\gamma)(3-\gamma)+\alpha(\alpha-\gamma)}
     {\alpha(3+\alpha-\gamma)}\,.
\end{eqnarray}
To illustrate, consider three cases.  For $\gamma > 0$ and
independent of $\alpha$, $c_0 \neq 0$ and
\begin{equation}
     n = \frac{6-\gamma}{2(2-\gamma)}\,,\;\;\;\;\;
     A=\frac{9-2\gamma}{4-\gamma} 
\end{equation}
recovering the result of Ref.\cite{qhs} and GS.
In the limit $\gamma \rightarrow 0$ and for $1 \leq \alpha < 2$, 
the first term in the expansion vanishes, $c_0 = 0$, and hence the 
second becomes the leading term. In this case one finds milder 
singularities,
\begin{equation}
     n = \frac{3-\alpha}{2}\,,\;\;\;\;\;
     A = \frac{9-\alpha}{4}  .
\end{equation}
Finally, in the limit $\gamma \rightarrow 0$ and 
$\alpha \rightarrow 2$, both $c_0$ and $c_1$ 
go to zero, the distribution function is non-singular, $n=0$,
and the density profile after the adiabatic growth of
the black hole has a $\gamma=3/2$ singularity, which is the
result obtained in Ref.~\cite{young} using an isothermal
(non-singular) distribution function.

A few comments on the results above. The derivation of the scaling in 
Eq.~(\ref{eq:scal}) verifies that the dark-matter spike is mainly formed by
particles which, before the black-hole growth, were very cold and close
to the Galactic center; i.e., particles with $E_i \rightarrow \Phi_i(0)$.
If the distribution function is singular in that limit (i.e., there are 
many such particles in the initial state), the spike is hugely
enhanced. As we have seen, this is possible even for a
dark-matter-density profile that is non-singular towards the
Galactic center
(but in this case the spike is always milder than for $\gamma \neq 0$).
Note also that the enhancement in the spike due to the singularity
in the distribution function is possible just if the black hole grows
exactly at the center of the initial dark-matter system.
If the black hole is formed away from the Galactic center,
the induced dark-matter spike will be mild. For a system with an 
isotropic velocity-dispersion tensor, it will be of the same
form we obtained above for a non-singular distribution function.

In the previous paragraph, we found for the NFW profile a spike
with $A=7/3$. For this result to hold, it is crucial to {\it postulate} 
that the initial density profile has a $1/r$ singularity all the way down
to the Galactic center. This behavior is just an extrapolation from an
N-body simulation with resolution not better than a kpc.
As the simulations show that larger structures form from the
merging of smaller substructures, such an extrapolation to scales
smaller than 10 pc or so may not be trivial.  More importantly,
even if the $1/r$ scaling is indeed preserved down to the inner
few pc in N-body simulations, it is unlikely to survive
the addition of baryons.  If the density profile
is assumed to be truncated in the innermost region,
we expect a shallower spike and a model-dependent suppression.
For example, if the truncation of a singular profile is performed introducing
a core radius, by replacing $\rho_i(r) \propto 1/r^{\gamma}$ with
$\rho_i(r) \propto 1/[1+(r/a)^{\alpha}]^{\gamma/\alpha}$, the spike
exponent $A$ will depend just on $\alpha$ (its expression can be
recovered from the analysis above substituting $\gamma = 0$ and
$\beta = \gamma$; note, however, that the normalization of the spike is
set by the other parameters as well).

Finally, the results in this Section show that the spike obtained
by assuming a single-power-law density profile ($\sim
1/r^{\gamma}$ for any $r$) with $\gamma\rightarrow 0$ will
probably not provide a good approximation to the true spike.
Most likely, the results in Ref.~\cite{gondolo}, where values of
$\gamma$ as low as $10^{-6}$ are considered, will be modified
assuming the more general double-power-law density profile of
the form in Eq.~(\ref{eq:prof1}).

\section{Instantaneous growth of the black hole}
\label{sec:inst}

We now consider what happens when the black hole is inserted
instantaneously to the center of the dark-matter distribution.
This will provide some indication of what happens to the
dark-matter distribution if the black-hole growth is too rapid
to be approximated as adiabatic.  We do this calculation under
the assumption that all particles are on circular orbits.

\begin{figure}[t]
\vskip 0.2cm
\centerline{\psfig{file=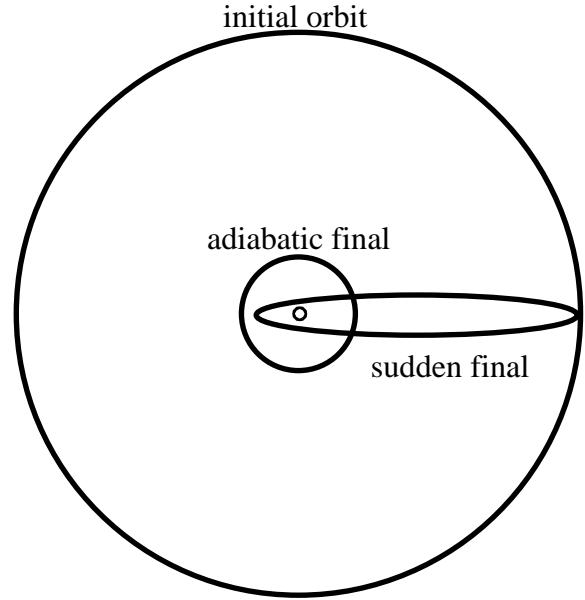,width=3.in}}
\vskip 0.5cm
\caption{Initial and final orbits for sudden and 
     adiabatic black-hole growth.  If the black-hole grows
     adiabatically, the initial large circular orbit becomes a
     final circular orbit of much smaller radius.  If the black
     hole appears suddenly, the initial circular orbit becomes
     an elliptical orbit shown (assuming that the final
     potential at these radii is dominated by the black hole).} 
\vskip -0.2cm
\label{fig:circles}
\end{figure}

We first provide a qualitative argument to show that the sudden black-hole
growth leads to a central dark-matter-density enhancement that
is not as dramatic as that for adiabatic growth.  
If the black hole grows adiabatically, the initial large
circular orbit becomes a final circular orbit of much smaller
radius.  Thus, the dark matter that was initially at a large
radius spends all of its time at a much smaller radius.  Now
suppose the black hole appears suddenly and dominates the
potential at these radii.  Again, the angular momentum of the
particle is conserved, but its energy changes in such a way that 
it follows an elliptical Keplerian orbit, 
as shown in Fig.~\ref{fig:circles}, in which its
largest radius is the initial radius.  Although the orbit reaches to
smaller radii than the circular orbit that arises from adiabatic 
growth, the velocity of the particle at these small radii is
large.  As a result, the particle on this orbit spends most of
its time at radii larger than the radius of the final orbit from 
adiabatic growth.

We now quantify these arguments.
When the black hole is added, the velocity of each particle
remains unchanged, but the gravitational potential is changed
suddenly.  The final profile is obtained by following each
particle along its subsequent orbit. The probability for a
particle that at the appearance of the black hole was at
galactocentric distance $r_0$ with a velocity $\vec{v}_0$ to be
between radii $r$ and $r+dr$ is
\begin{equation}
     P(r) dr = \frac{dt}{T(r_0,\vec{v}_0)} = 
     \frac{1}{T(r_0,\vec{v}_0)} \frac{dr}{v_r(r,r_0,\vec{v}_0)},
\end{equation}
where $T$ is half of the period of the particle, and $v_r$ is the 
radial component of its velocity.

If all particles are on circular orbits, then the calculation is 
simplified, as the initial radial velocities are all zero.
We thus need to integrate the probability that a particle that
is observed at radius $r$ in the final configuration came from
an orbit at radius $r_0$ in the initial configuration.
Doing so, we find that the final radial profile is given by:
\begin{equation}
     \rho_f(r) = \frac{1}{r^2} \int_{r_{\rm min}}^{r_{\rm max}} dr_0
     r_0^2 \rho_i(r_0) \frac{1}{T(r_0)} \frac{1}{v_r(r,r_0)},
\end{equation}
where
\begin{equation}
     v_r = \sqrt{2 \left[\Phi_f(r_0)-\Phi_f(r)\right]
     + v_c^2(r_0) \left(1-r_0^2/r^2\right)},
\end{equation}
with $v_c$ the circular velocity in the initial state.
The extremes of integration $r_{\rm min}$ and $r_{\rm max}$ are 
the minimum and maximum initial radii for particles contributing
to the final profile at the radius $r$.  For such radii,
$v_r(r,r_{\rm min}) = v_r(r,r_{\rm max}) = 0$.  From the expression for
$v_r$, it follows that $r_{\rm min} = r$; i.e., a particle on a circular
orbit shrinks in general to a rosette orbit with apocenter equal
to the radius of the initial orbit.

We compute $\rho_f$ for the initial NFW profile
introduced in Section~\ref{sec:anal}. Again, we assume
$\Phi_f(r) \cong \Phi_f(r) - G M_{\rm BH}/r$; the result is shown as
a dash-dot curve in Fig.~\ref{fig:fig1}. 
Although the black hole steepens the profile, the steepening is
not nearly as dramatic as if the black hole is introduced
adiabatically.

The final profile has a $r^{-4/3}$ singularity, a result that
can be understood analytically.
In the small-$r$ limit, the potential is dominated by the black
hole, and $T(r_0) \propto r_0^{3/2}$, while $v_r$ becomes
\begin{eqnarray}
     v_r & = & \sqrt{\frac{2 G M_{\rm BH}}{r r_0}} \nonumber \\
     && \times \sqrt{(r_0-r)\left(1-\frac{M_i^{h}(r_0)}{2 M_{\rm BH}} 
     \left(1+\frac{r_0}{r}\right)\right)}\,.
\end{eqnarray}
For the NFW profile, the halo mass $M_i^h(r_0)$ enclosed within the
radius $r_0$ is proportional to $r_0^2$, $r_{\rm max}$ can be
computed exactly, and the final profile is
\begin{eqnarray}
     \lefteqn{\rho_f(r) \sim \frac{1}{r} \int_{r}^{r_{\rm max}} dr_0
     \frac{1}{\sqrt{(r_0-r)(r_{\rm max}-r_0)}}} \nonumber \\ 
     && \times\frac{1}{\sqrt{r_0^2-(r r_0)/(k r_{\rm max}^2) - r/(k
     r_{\rm max})}}
     \sim \frac{1}{r^{4/3}},
\end{eqnarray}
as found numerically and shown in Fig. \ref{fig:fig1}.

Heuristically, our circular-orbit model should generalize to any orbit 
with a finite peri-to-apocenter ratio. The apocenter distribution is 
unchanged after the implanting of the black hole, and the pericenter is
determined by the angular-momentum distribution, which is a
power law differing from that of the circular orbits by a
scaling.  So we expect the pericenter to move in according to
the same power law as for circular orbits and thus produce a
density power law as before except for normalization.

\section{Models with off-centered Black Hole}
\label{sec:spiral}

The black hole, or at least its seed, might not form exactly at the
center of the dark-matter halo due to deviations from spherical 
symmetry in various galaxy formation processes.
If the black hole forms off-center, then it will slowly sink via 
dynamical friction to the center of the Galactic potential well.
We sketch in this Section the number of competing effects
entering this problem.

\subsection{Enough time to spiral in?}

\begin{figure}[t]
\vskip -1.0cm
\centerline{\psfig{file=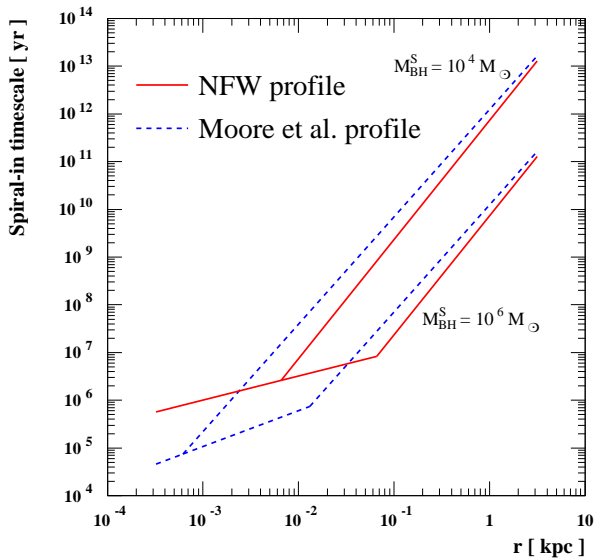,width=3.5in}}
\caption{Predicted time for a seed black hole of mass 
$M_{\rm BH}^S=10^4 \msun$ and $10^6 \msun$ to spiral in to the Galactic
center from a radius $r$ in an
initially pure NFW or pure Moore halo profile (no baryons). It takes
longer than a Hubble time for a $10^4 \msun$ 
black hole to spiral from $r \gg 100$~pc, and for a $10^6 \msun$ black hole 
to spiral from $r \gg 1$~kpc.  The spiral-in time is estimated
with Chandrasekhar's formula for dynamical friction,
depending slightly on the Coulomb factor and 
the eccentricity of the orbit of the black hole.
Note the curves join together after a break at very small radius 
when the enclosed ambient mass becomes comparable to 
the mass of the infalling hole.
Inside this radius, the hole would fall in quickly on a
dynamical time scale, which is independent of $M_{\rm BH}^S$. 
}
\vskip -0.2cm
\label{fig:rece}
\end{figure}

The process of dynamical friction
is not efficient enough to center the black hole regardless of its
initial mass and of its initial distance from the center of mass.
The spiral-in
timescale for a very small black hole or for a black hole born very
far from the center of the Galaxy, may be longer than the age of the Galaxy
itself. For a black-hole seed of mass $10^4 \msun$, the timescale
to spiral in the potential well of the NFW profile we considered
is roughly comparable to the Hubble time if its initial location
is 100~pc away from the center of the dark-matter system. This is 
illustrated in Fig~\ref{fig:rece}.  Also shown is the case of a 
black-hole seed of mass $10^6 \msun$, for which the timescale is
of the order of 10~Gyr if it spirals in from 1~kpc.  The conclusion is that
if, as plausible, the black-hole seed is 
generated through some violent process off-center, it will not
get to the center until almost fully formed. 

\subsection{Merger and growth at the center}

What happens when the seed BH arrives at the center?  There are
two competing effects with the dark-matter distribution near the 
Galactic center: (a) it may be smoothed out by the time-dependent perturbation
induced by the incoming black-hole; (b) it may be squeezed in by subsequent 
at-the-center growth of the black hole.

Effect~(a) has been investigated with N-body simulations in 
Ref.~\cite{NakanoMakino}. It was shown there that the back-reaction to
the spiral-in of the black hole leads to the formation of a weak density
cusp, with an $r^{-0.5}$ singularity, regardless of the initial 
density profile.  More precisely, the same features are obtained 
in simulations with the King model (which has a core radius)
and with the Hernquist profile (which is singular at the Galactic 
center and given by Eq.~(\ref{eq:prof1}) with
$(\alpha,\beta,\gamma)=(1,4,1)$). Furthermore, the size of the
weak-cusp region is 
found to be correlated to the black-hole mass, with roughly the 
total mass of the region affected by the black hole equal to the 
black-hole mass. The proposed explanation for this result is as 
follows: The sinking of the black hole induces a heating process 
that leaves no particle in the system with energy below some 
minimum energy $E_0$. We saw above in the case of adiabatic growth
at the center of the halo that the spike in the final density
profile was due to initially cold particles. 
If such particles are removed from the system, neglecting the dependence 
on angular momentum in the distribution function, one finds that
the final density profile goes at small radii like~\cite{NakanoMakino}
\begin{eqnarray}
     \rho_f(r) &=& 4\pi \int_{\Phi_f(r)}^0 f_f(E_f) 
     \sqrt{2\left[E_f-\Phi_f(r)\right]} \nonumber \\
     &=& 4\pi \int_{E_0}^0 f_f(E_f) 
     \sqrt{2\left[E_f-\Phi_f(r)\right]} \nonumber \\
     &\sim& \sqrt{-\Phi_f(r)} \sim \sqrt{\frac{1}{r}}\,.
\end{eqnarray}

We implement this result in our picture, with a simple scheme.
We suppose we have an initial dark-halo profile of the form
in Eq.~(\ref{eq:prof1}). Then we postulate the off-center
formation of a black-hole seed of mass $M_{\rm BH}^S$, which 
we keep as a free parameter.
Since we are supposing that the black-hole seed appears off-center,
regardless of whether it forms quickly or slowly, it induces a 
very mild dark-matter spike around it; we neglect this effect.
We model then the modification to the dark-halo profile due to 
the sinking of the black-hole seed in the center of the Galaxy 
by replacing the initial density profile with the following:
\begin{eqnarray}
     \rho_{\rm int}(r) &=& \rho_0 \left(\frac{R_0}{r}\right)^{0.5}
     \left[\frac{1+(R_0/r_s)}{1+(r/r_s)}\right]^{\gamma-0.5} 
     \nonumber \\
     &&\times \left[\frac{1+(R_0/a)^{\alpha}}{1+(r/a)^{\alpha}}\right]^
     {(\beta-\gamma)/\alpha},
\label{eq:prof2}
\end{eqnarray}
which is essentially a modified Zhao model with a break at both $r_s$ and $a$.
The new length scale $r_s$ we introduce is determined by requiring
that the mass of the dark halo within $r_s$ to be equal to 
$M_{\rm BH}^S$. 

Moving onto Effect~(b), we model the accretion of matter onto the black hole
so that $M_{\rm BH}^S$ grows to the mass 
we now observe $M_{\rm BH}$ as a slow adiabatic
process, and compute the final dark-matter density,
treating the adiabatic contraction in the circular-orbit
approximation described above.

\begin{figure}[t]
\vskip -1.0cm
\centerline{\psfig{file=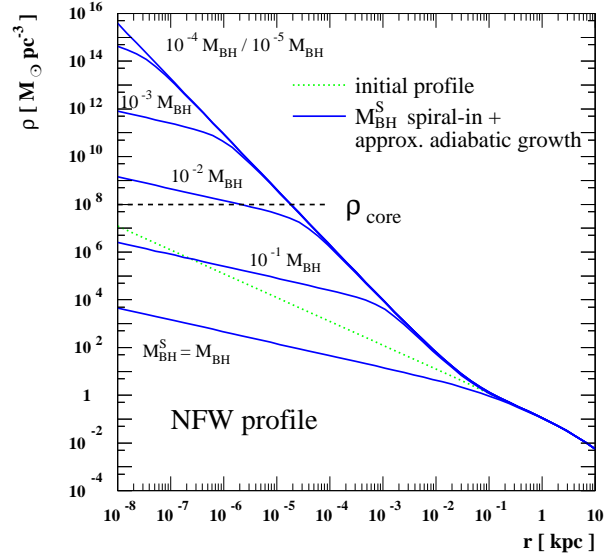,width=3.5in}}
\caption{Modification of an NFW dark-matter-density profile due to the
off-center formation of a black-hole seed of mass $M_{\rm BH}^S$, its spiral in
the center of the dark-matter system and its adiabatic growth to
the present-day mass of the black hole at the Galactic center. 
The cases for a few different values for the black-hole-seed mass are 
plotted. $\rho_{\rm{core}}$ is the maximum WIMP density
above which WIMPs are depleted by pair annihilations.
} 
\vskip -0.2cm
\label{fig:fig2}
\end{figure}

In Fig.~\ref{fig:fig2} we plot $\rho_f$ for the initial 
NFW profile introduced in Section~\ref{sec:anal}
and for a few choices for the mass of the black-hole seed.
As can be seen, if the black hole that spirals in has a mass 
comparable to (more than a tenth of) 
the mass we observe now at the center of the Galaxy, the dark-matter density
close to the black hole is, in this picture, lower than the 
density in the initial profile. This indicates that the number
of dark-matter particles expelled from the inner Galaxy by the
black-hole seed is larger than the number of those 
attracted later on by the adiabatic deepening of the potential
well due to the increase in mass of the black hole.

The adiabatic contraction process dominates for $M_{\rm BH}^S \lta M_{\rm BH}/10$.  
For $M_{\rm BH}^S \lta M_{\rm BH}/200$, the end result is a dramatic
enhancement in the dark-matter density around the black hole.
Although the final density profile
obtained with this procedure differs from what we found in case
of adiabatic growth of the black hole at the center of the
dark-matter system, the difference occurs only at densities
higher than the maximum WIMP density $\rho_{\rm{core}}$.  Thus,
if a black-hole seed with less than 1/200th the mass of
the final black hole is let to grow adiabatically at the center, then a spike
that is essentially indistinguishable accordance with the GS
result could be formed.  The trouble is that such small seed cannot
spiral in the center quickly enough, as shown by Fig.~\ref{fig:rece}.
So the conclusion
is that if the GS spike is to grow, the black hole
must form within the inner 50 pc of the dark-matter
distribution:  If it forms outside this radius with a large
mass, it will destroy the cusp as it spirals in, and if it forms 
well outside this radius with a small mass, it will not have
enough time to spiral in.

Although we have considered so far just the NFW profile,
the procedures we outlined can be applied to any density
profile. As another example, we consider the dark-matter-density
profile obtained in the high-resolution N-body simulation of
Moore et al.~\cite{moore}. In this case, the initial dark-matter
profile is more cuspy towards the Galactic center and is
given by Eq.~(\ref{eq:prof1}) with
$(\alpha,\beta,\gamma)=(1.5,3,1.5)$ (hereafter the Moore et al. profile). 
Again we make a choice for the values of our galactocentric 
distance, the local halo density, and the length scale in agreement 
with available dynamical constraints: $R_0 = 8$~kpc,
$\rho_0 = 0.3$~GeV~cm$^{-3}$, and $a = 28$~kpc.

\begin{figure}[t]
\vskip -1.0cm
\centerline{\psfig{file=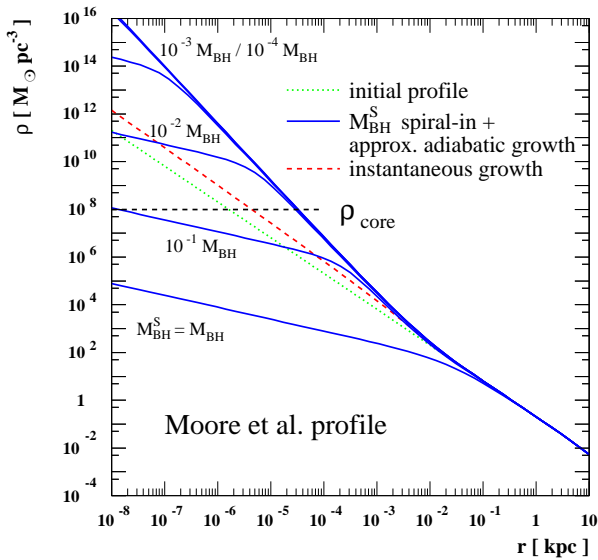,width=3.5in}}
\caption{The same as in Fig.~\ref{fig:fig2}, but for a Moore et
al.\ dark-matter-density profile. Also shown is the modification 
of the profile for sudden growth of the black hole at the center
of the dark-matter system (dashed curve).} 
\vskip -0.2cm
\label{fig:fig3}
\end{figure}

Results in this case are shown in Fig.~\ref{fig:fig3}.
The initial profile is shown as a dotted curve; the dashed curve
shows the slight increase in density in case of instantaneous
growth of the black hole according to the treatment in 
Section~\ref{sec:inst}. The solid curves give the final profile
in the case described above in this Section.  We see again that
for large values of the mass of the black-hole seed, the net 
effect is a decrease in the dark-matter density. For values
$M_{\rm BH}^S \lta M_{\rm BH}/50$, the enhancement in the final density
reproduces, for $\rho_f < \rho_{\rm{core}}$, the result one gets
just assuming the adiabatic growth at the center of the dark 
matter system. The timescales to re-center black hole seeds of
$10^4 \msun$ and $10^6 \msun$ are again shown in Fig~\ref{fig:rece}, 
and the same discussion we outlined for the NFW case applies here.

\section{Taking stars into account}
\label{sec:stars}

So far we have used a schematic picture of the Galaxy which 
included just dark matter and the black hole, neglecting
other baryonic components.  More realistically, the stellar
components may outweigh the dark matter in the inner part of the 
galaxy in the initial configuration before the black hole grows.
We treat this problem here.

In the standard sequence of galaxy formation, a roughly
spherical overdensity, in which the baryons and dark matter are
distributed homogeneously, undergoes gravitational collapse.  
The collisionless matter then relaxes to an NFW or Moore
profile, as indicated by N-body simulations, while the baryonic
matter shocks, heats, and then cools to a disk and/or bulge in
the gravitational potential well generated by both the dark and
the baryonic matter.  The radiative cooling and collapse of
baryons is likely to be a slow process and hence it should be a
good approximation to treat it as an adiabatic contraction of
the baryons.

We sketch this process in one example. We reconsider again the case of
off-center formation discussed in the previous Section, refining the
definition of initial and final states. In the initial state we add the
extra component due to baryons (same density profile as for dark matter
but scaled down of a factor $\Omega_b/\Omega_M$). In the final state
we add the stellar bulge and disk.  We pick standard forms for them, 
compatible with available observations in the outer part of the Galaxy
and with the rotation curve close to the black hole (we do not reproduce
the functional forms here as they do not enter critically in the 
discussion below). The spiral-in of the black-hole seed is treated in
the same way as in Section~\ref{sec:spiral}.
The adiabatic contraction is performed using the circular-orbit 
approximation. We also replace the disk with a spherical mass distribution
that produces the same rotation curve.  This simplification allows us
to still treat the problem as spherically symmetric and does not enter
crucially in the results we focus on.

\begin{figure}[t]
\vskip -1.0cm
\centerline{\psfig{file=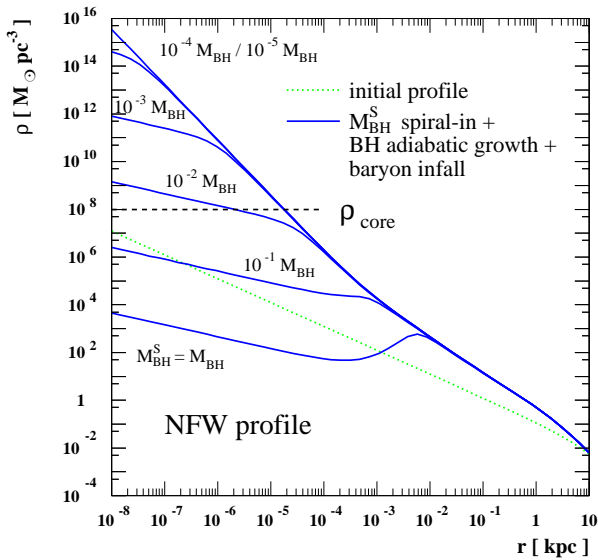,width=3.5in}}
\caption{The same as in Fig.~\ref{fig:fig2}, but taking into account 
the adiabatic contraction of baryons that form the bulge and the disc}
\vskip -0.2cm
\label{fig:fig4}
\end{figure}

Results are shown in Fig.~\ref{fig:fig4}.  The adiabatic
contraction of the baryons in the bulge and the disk induces a significant
enhancement of the dark-matter density at intermediate galactocentric 
distances (few pc up to few kpc). In the innermost region,
however, the main role is still played by the black hole.

\section{Conclusions}
\label{sec:concl}

We have tried to determine
the initial conditions required for growth of this spike by
considering several alternative formation scenarios.  We have
found that for the spike to arise, the majority of the
black-hole mass must grow on a timescale long compared with
$10^7$ years and within 50 pc of the center of a cuspy dark-matter
distribution.  Moreover, the dark-matter particles that
make up this spike must be initially in very cold
orbits near the Galactic center.
A spike essentially equivalent to the GS spike could in
principle arise if a low-mass black hole spiraled in to the
Galactic center and then grew in mass, but the spiral-in
timescale for such low-mass black holes is larger than the
Hubble time.

If we understood precisely how the black hole in the Galactic
center formed, we could say more definitively whether the GS
spike arises or not.  Unfortunately, however, the details of
black-hole formation are far from clear.  The present-day growth 
rate for the Milky Way's black hole as well as the peak of AGN
activity at redshifts $z\sim2$ suggest that a good fraction of
the black-hole mass was in place at early times.
The most promising mechanisms for black-hole production 
at high redshift do not seem to contain the ingredients required 
for the GS spike (see, e.g., Refs. \cite{bhs}).  N-body
simulations show that at that time,
dark halos are forming via rapid and often violent mergers.
Moreover,
hydrodynamic simulations show that when two galaxies collide,
they can leave behind a cold dense baryonic gas that can then
collapse to form a supermassive black hole.  The rate for
black-hole growth is estimated to be orders of magnitude larger
than the current rate (see, e.g., Ref.~\cite{Zhaoetal}),
although a timescale long compared with $10^7$ years, the
period for the orbits that eventually wind up in the spike, is still
conceivable.  However, even if the timescale is reasonable,
the dark halos for each of the two initial galaxies pass through 
each other almost undisturbed and then later relax to
the dark-matter profile for a single halo.  The timescale for
this relaxation should be even
longer.  Moreover, it is hard to believe that the cold orbits
required for the GS spike would be found right at the center of
the dark-matter distribution so soon after this very violent
collision.  The mass density in these inner regions will be
dominated by the baryons, and baryons can shock, form molecular
clouds, stars, star clusters, and supernova-driven winds.  The
dynamical effects of all these processes are heuristically
expected to disrupt cold orbits.  Finally, it is not clear why
the centers of the dark-matter and gas distributions should be so
coincident in this violent formation.  (Although the black hole
in our Galaxy coincides fairly well at the current epoch with
what appears to be the dynamic center of the Galaxy, this is not 
necessarily the case in other galaxies, such as M31.)

An alternative (although not mutually exclusive) possibility is
that a $10^3-10^5 \msun$ black hole could form at a redshift
$z\sim30$ when molecular-hydrogen cooling permits the collapse of the first
baryonic objects.  The remaining $10^6 \, \msun$ of mass could
then be accreted during the subsequent generations of mergers
that eventually give rise to a galactic halo.  Although it would 
not be unusual for the initial high-density peak in the
primordial mass distribution that gave rise to this black-hole
seed to be near the center of the resulting galactic halo, it
would be somewhat surprising if it was {\it so} close to the
center.

In conclusion, we have determined the initial conditions for the
growth of the GS spike and given several reasons to doubt they
were manifest in our own Galaxy.  Given these doubts, null
searches for annihilation radiation from the Galactic center
should not be interpreted as evidence against WIMP dark matter,
nor against a cuspy halo.  \comment{ With the advent of a variety of new
observational probes of stellar dynamics near extragalactic
supermassive black holes in the Universe today, and of the
high-redshift Universe, where these black holes were born, we
should soon be able to piece together a more precise recipe for
black-hole formation.  Until then, however, the precise
distribution of dark matter near the Milky Way's supermassive
black hole will remain uncertain.}

\acknowledgments
We thank A. Loeb for useful discussions.  This work was supported in
part at Caltech by NSF AST-0096023, NASA NAG5-8506, and DoE
DE-FG03-92-ER40701, and at Trieste by the RTN project under grant
HPRN-CT-2000-00152. H.S.Z. wishes to thank Institut d'Astrophysique de
Paris and Sterrewacht Leiden for financial supports and Caltech for
hospitality during his visit.


{}

\end{document}